\documentclass[12pt]{article}
\usepackage{amsfonts}
\usepackage{amssymb}
\usepackage{amscd}
\usepackage{amsmath}

\usepackage[usenames]{color}

\newcommand{\BEQ}{\begin{equation}}
\newcommand{\EEQ}{\end{equation}}
\def\bea{\begin{eqnarray}}
\def\eea{\end{eqnarray}}
\def\nn{\nonumber}

\newtheorem{Th}{Theorem}
\newtheorem{Lem}{Lemma}

\newtheorem{Def}{Definition}

\newtheorem{Rem}{Remark}

\def\bea{\begin{eqnarray}}
\def\eea{\end{eqnarray}}

\def\bes{\begin{equation*} \begin{split}}
\def\ees{\end{split} \end{equation*}}

\def\C{{\mathbb{ C}}}

\def\Z{{\mathbb{ Z}}}

\usepackage{graphicx}
\graphicspath{{pics/}}

\begin{document}

\title{Asymmetric Hopfield neural network and \\Twisted tetrahedron equation}

\author{\rule{0pt}{7mm} Dmitry V. Talalaev\thanks{dtalalaev@yandex.ru}\\
{\small\it
Moscow State University, Faculty of Mechanics and Mathematics}\\
{\small\it 119991 Moscow, Russian Federation}}

\date{}

\maketitle

\abstract{
We generalize the approach of \cite{T1} for the case of the Hopfield neural network in the recall stage on a triangular lattice with isotropic weights. It appears that some properties of this model, in particular the probability of passing a trajectory in time dynamics, obeys the Gibbs distribution with a partition function having a vertex realization. Moreover the corresponding weight matrix satisfies the TTE - some deformation of the Zamolodchikov tetrahedron equation, the latter playing the role analogous to the Yang-Baxter equation in 3-dimensional statistical models.
 }

\vskip 5mm
~\\
\tableofcontents
\vskip 2cm
\section{Introduction}
Models of artificial and natural neural networks for a long time have been shown to be related to the integrable models in lattice statistical physics. The main emphasis of this work is on some new kind of relation between the Ising model and the Hopfield model of associative memory.

\subsection*{Acknowledgments} 
The work was partially supported by the RFBR grant 17-01-00366.

\subsection{Hopfield model}
The Hopfield network \cite{Hop} is one of the fundamental models of content-addressable associative memory system. This model have demonstrated some very interesting collective properties of the neural network behavior and is used in artificial neural networks as like as in neurophysiology \cite{Tso} in research of memory capacities, memory retrieval process, short-term plasticity, working memory properties and many other questions. The model is represented by the complete graph with $N$ vertexes (neurons) with a connectivity matrix $W_{ij}$ characterizing the conductivity of the synapse between $i$-th and $j$-th neurons. At each time the system is characterized by its neurons states $\{x_i\},i=1,\ldots,N$ $x_i=\pm 1.$ Our interest is focused on the network which undergoes the synchronous evolution in discrete time. In the deterministic version the state of $i$-th neuron at the next step $x'_i$ is determined by the formula:
\bea
x'_i=\left\{
\begin{array}{ccc}
1 & \mbox{if}  & \sum_{j} W_{ij}x_j >t_i \\
-1 & \mbox{if} &  \sum_{j} W_{ij}x_j < t_i \\
x_i & \mbox{else.} &
\end{array}\right.
\eea   
We principally deal with the probabilistic Hopfield model in which the transition is performed with the probability specified by the Fermi sigmoid function
\bea
\label{prob}
P(x',x)=\prod_{i}(1+e^{-\beta x'_i(\sum_j W_{ij}x_j-t_i)})^{-1}.\nn
\eea
For the threshold level $t_i=0$ this expression can be rewritten as 
\bea
P(x',x)= {e^{-\frac \beta 2 \sum_{ij}W_{ij}x'_i x_j}}/ {\sum_{x''}e^{-\frac \beta 2 \sum_{ij}W_{ij}x''_i x_j}}.\nn
\eea
The similarity of such an expression with the partition function of the Ising model was remarked in \cite{Hop}. We make this resemblance more formal in the next section. 

We need to say some words about the learning rules. Following the Hebbian paradigm the learning of the Hopfield network on the set of $n$ patterns $\{\epsilon^1,\ldots,\epsilon^n\}$ is given by the formula
\bea
W_{ij}=\frac 1 n \sum_{k}\epsilon_i^k \epsilon_j^k.\nn
\eea
In what follows we start with the learned network with isotropic weight function.
The work of the network in the recall stage consists of iterative time dynamics defined by the transformation probability (\ref{prob}). The principal question in this domain is the basins of attraction for such a process.

The principal interplay of the Hopfield model and the Ising one is due to the energy argument. It turns our that for the symmetric weight function the expression
\bea
E=-\frac 1 2 \sum_{i,j} W_{ij} x_i x_j -\sum_i t_i x_i\nn
\eea
plays the role of the energy in the sense that in time dynamics it either lower or stays the same. This observation allows to analyze the asymptotic behavior of the Hopfield model. The stable points for the Hopfield time dynamics are hence the states of local minimum energy for the Ising model on the same lattice. In this paper we explore another relationship between two such models: the Hopfield model on a two-dimensional lattice and the Ising model on the 3-dimensional lattice, making one of the spacial directions relevant to the time evolution.

\subsection{Time evolution and the Ising model}
\begin{figure}[h!]
\center
\includegraphics[width=80mm]{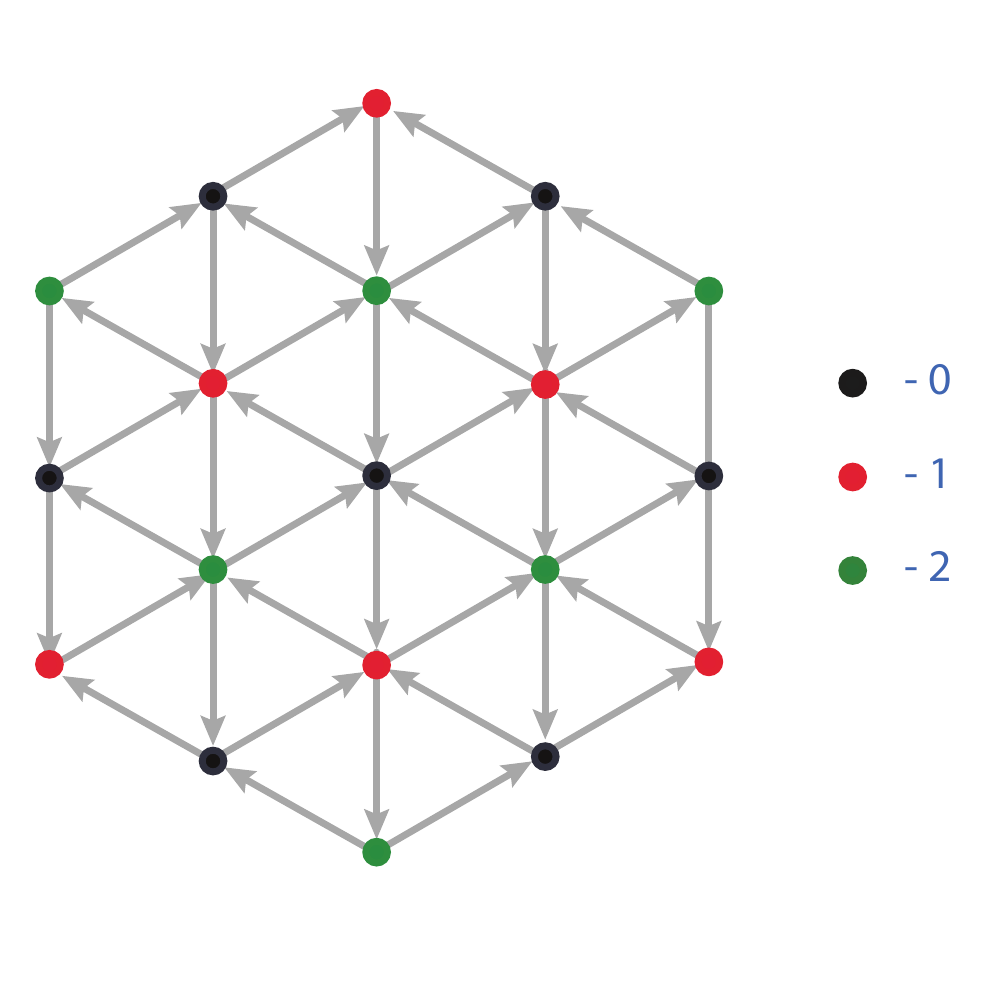}
\caption{Triangular lattice}
\label{pic-lattice}
\end{figure}

Let us consider the Hopfield model on the triangular lattice (figure \ref{pic-lattice}) colored by 3 colors in $\Z/3\Z.$ The weights are nontrivial for neighbors with appropriate colors: for the neuron with the color $c$ we choose only such influencing neurons whose colors are $(c-1) ~mod ~3.$ This network is not symmetric despite the most traditional normalization.

This lattice corresponds to the projection of the cubic lattice to the plane $i+j+k=0$. The vertexes with different colors represent the planes $i+j+k=c~ mod~ 3.$ The temporal behavior of this lattice is equivalent to the 3 independent 3-dimensional cubic lattices. We then consider the one of these 3 lattices.

The conditional probability that the model passes throw the states with free initialization data is:
\bea
P=\prod_{i+j+k=a}^b\frac {e^{-\beta s_{ijk}(s_{ijk-1}+s_{i-1 jk}+s_{i j-1k})}}{ch(\beta (s_{ijk-1}+s_{i-1 jk}+s_{i j-1k}))}\nn
\eea
\begin{Lem}
Let $\sigma_2$ be the second symmetric polynomial on 3 variables $s_1,~s_2$ and $s_3$ taking values $\pm 1.$ Then
\bea
\label{eq1}
ch(\beta(s_1+s_2+s_3))=ch^3\beta+sh^2\beta ch \beta \sigma_2.
\eea
\end{Lem}
Demonstration is straightforward.
\\
Moreover we have:
\bea
\label{eq2}
e^{\gamma \sigma_2}=(ch^3\gamma +sh^3 \gamma)+(ch^2 \gamma sh \gamma+ch \gamma sh^2 \gamma)\sigma_2.
\eea
Expressions (\ref{eq1}) and (\ref{eq2}) turn out to be proportional if 
\bea
\label{eq3}
th^2 \beta=th \gamma/(1- th \gamma+th^2 \gamma)
\eea
With $\gamma$ solving the equation (\ref{eq3}) the conditional probability for the Hopfield model is identical to the Gibbs distribution for the anisotropic Ising model on cubic lattice with additional diagonals with specially chosen anisotropy parameters and with border terms
\bea
P(s)=\frac 1 Z \prod_{i,j,k} e^{-\beta s_{ijk}(s_{i-1 j k}+s_{i j-1 k}+s_{i j k-1})-\gamma(s_{i-1 j k}s_{i j-1 k}+
s_{i j-1 k} s_{i j k-1}+s_{i j k-1} s_{i-1 j k})}.\nn
\eea
\begin{figure}[h!]
\center
\includegraphics[width=70mm]{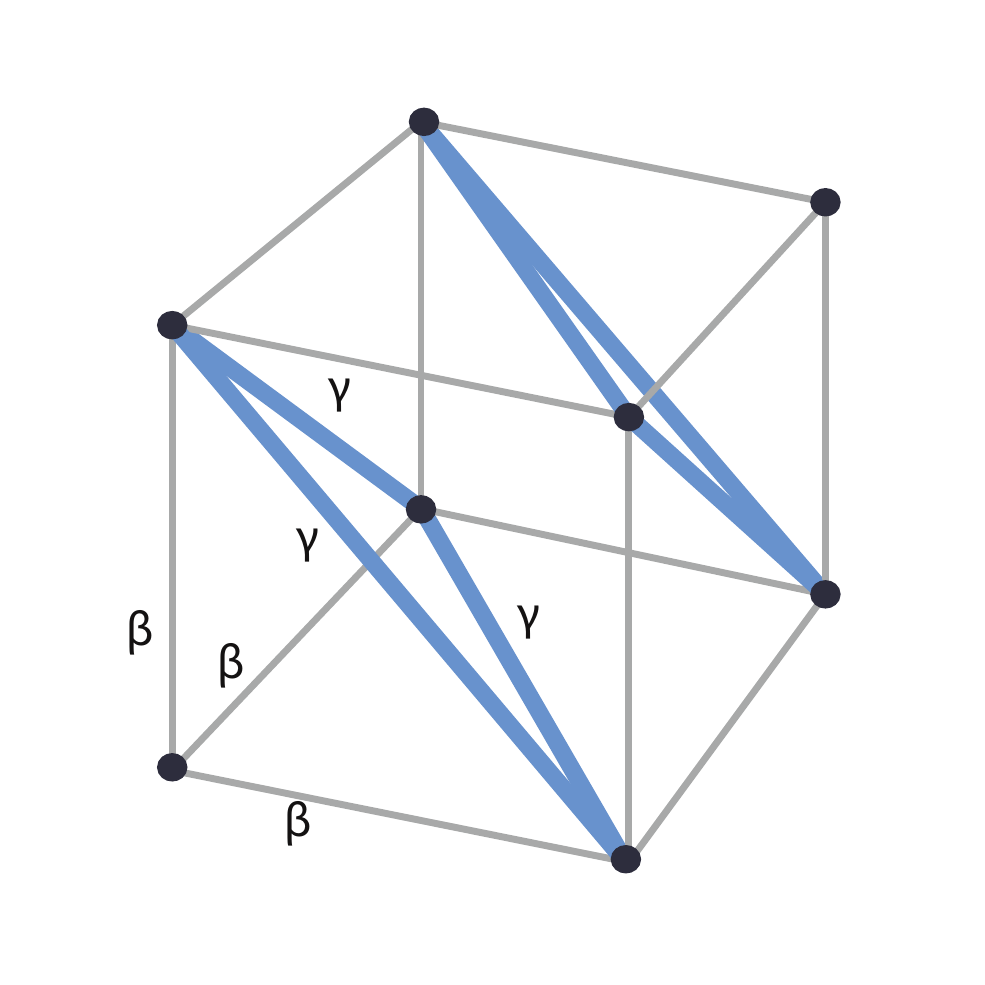}
\caption{Ising model lattice}
\label{edges}
\end{figure}

The main problems stated in the Hopfield theory are related with stationary points of this dynamics and with correlations. For example the correlation length in the Ising model and its particular dependence on both parameters $\beta$ and $\lambda$ is related with such questions as cluster dimension and time evolution.

\subsection{3D Ising model and TTE}
Let us remind the main steps of the construction of \cite{T1}.

\paragraph{Dual lattice}
The isotropic Ising model is described by the Hamiltonian 
\bea
H(\sigma)=\sum_{d(i,j)=1} \sigma_i \sigma_j\nn
\eea
where $i\in\Lambda$ is a periodic 3d lattice, $\sigma_i$ is the spin variable associated to the $i$-th vertex. $d(i,j)$ is the standard (Manhattan) metric on the cubic lattice.
The partition function is defined by the Gibbs rule:
\bea
\label{part}
Z(t)=\sum_{(\sigma)}\exp(t H(\sigma))
\eea
where the sum is taken over the space of all spin configurations.

To work with the dual lattice we introduce the variables $s_{ij}=\sigma_i \sigma_j$ on edges and associate the space $\C^2$ to each edge of $\Lambda.$ Then we consider the dual lattice $\Lambda^*$ those vertices are 3-cubes of $\Lambda$, the edges of $\Lambda^*$ are 2-faces of $\Lambda.$ We associate a vector space $V_f=(\C^2)^{\otimes 4}\simeq \C^{16}$ to each edge of the dual lattice $\Lambda^*.$ 

\paragraph{Recursion on the moduli of solutions for the $n$-simplex equation}
In \cite{KST} we define a moduli space $\mathfrak{S}_n(X)$ of solutions for the set theoretical $n$-simplex equation in correspondences with the underlying set $X$. In \cite{T1} we constructed a recursion
\bea 
\tau: \mathfrak{S}_n(X)\to \mathfrak{S}_{n+1}(X^{2n})\nn
\eea
assigning a set of colors of 1-edges of a 2-face to this 2-face. We then applied such a recursion to the solution for the  Yang-Baxter equation given by the matrix
\bea
R_M=\left(\begin{array}{cccc}
1 & 0 & 0 & 1\\
0 & 1 & 1 & 0\\
0 & 1 & 1 & 0\\
1 & 0 & 0 & 1
\end{array}\right).
\eea
\begin{Rem}
This matrix encodes the conditions on edge-spins for the Ising model.
\end{Rem}
Then $\tau(R)$ produces an auxiliary weight matrix $W_0$ for the 3D Ising model.

\paragraph{Counting edges of the 3-cubes}
First let us make some remarks on the notations in hypercube combinatorics.
The $n$-cube subfaces are denoted by sequences $(a_1,\ldots,a_n)$ where $a_i$ are either $1$ or $0$ or $*$. The number of stars is the dimension of a subface.
Then define a sequence
\bea
\zeta=(0,1,\ldots,).\nn
\eea
\begin{Def}
An $(n-1)$-subface of an $n$-face is called incoming if the $i$-th $*$ is replaced by $\zeta_i,$ otherwise it is called outgoing.
\end{Def}
\begin{Rem}
We also introduce a lexicographic order on subfaces. For example the standard $2$-cube has $4$-edges, 2 incoming and 2 outgoing, we denoted their orders in the second line of the table.
\begin{table}[h!]
\begin{center}
\begin{tabular}{|c|c|c|c|}
\hline
0*& *1 & 1*& *0\\
\hline
$i_1$ & $i_2$ & $o_1$ & $o_2$ \\
\hline
\end{tabular}
\end{center}
\end{table} 
\end{Rem}

We then choose a cocycle on 3-cube colorings determined by the following choice of 3 edges on each 3-cube of a 4-cube given by the  tables
\begin{table}[h!]
\caption{Left side of the TE}
\label{tab-LTE}
\begin{center}
\begin{tabular}{|c|c|c|c|}
\hline
0***& *1** & **0* & ***1\\
\hline
01*0 & 011* & 0*00 & 00*1 \\
000* & *100 & 110* & *111 \\
0*11 & 11*1 & *001 & 1*01 \\
\hline
\end{tabular}
\end{center}
\end{table}

\begin{figure}[h!]
\center
\includegraphics[width=80mm]{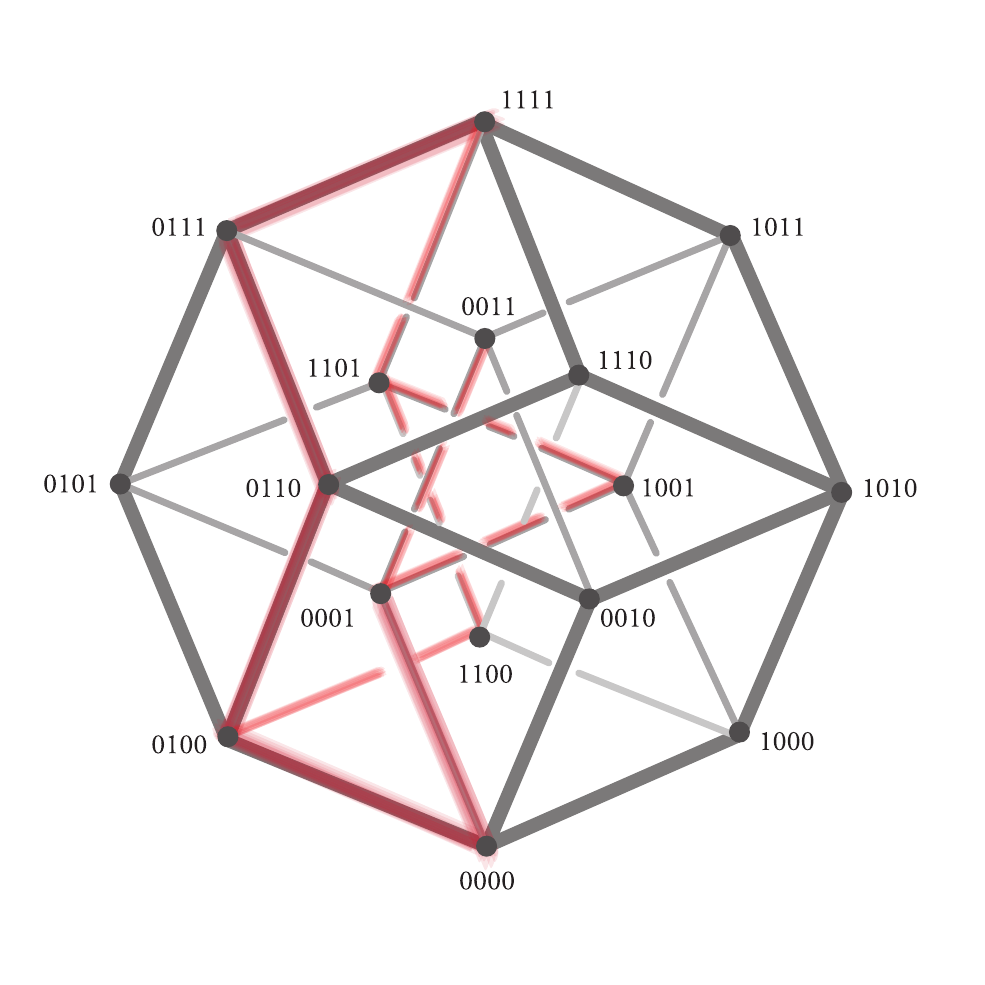}
\caption{Chosen subgraph in the 4-cube}
\label{pic-edges}
\end{figure}

\begin{figure}[h!]
\center
\includegraphics[width=80mm]{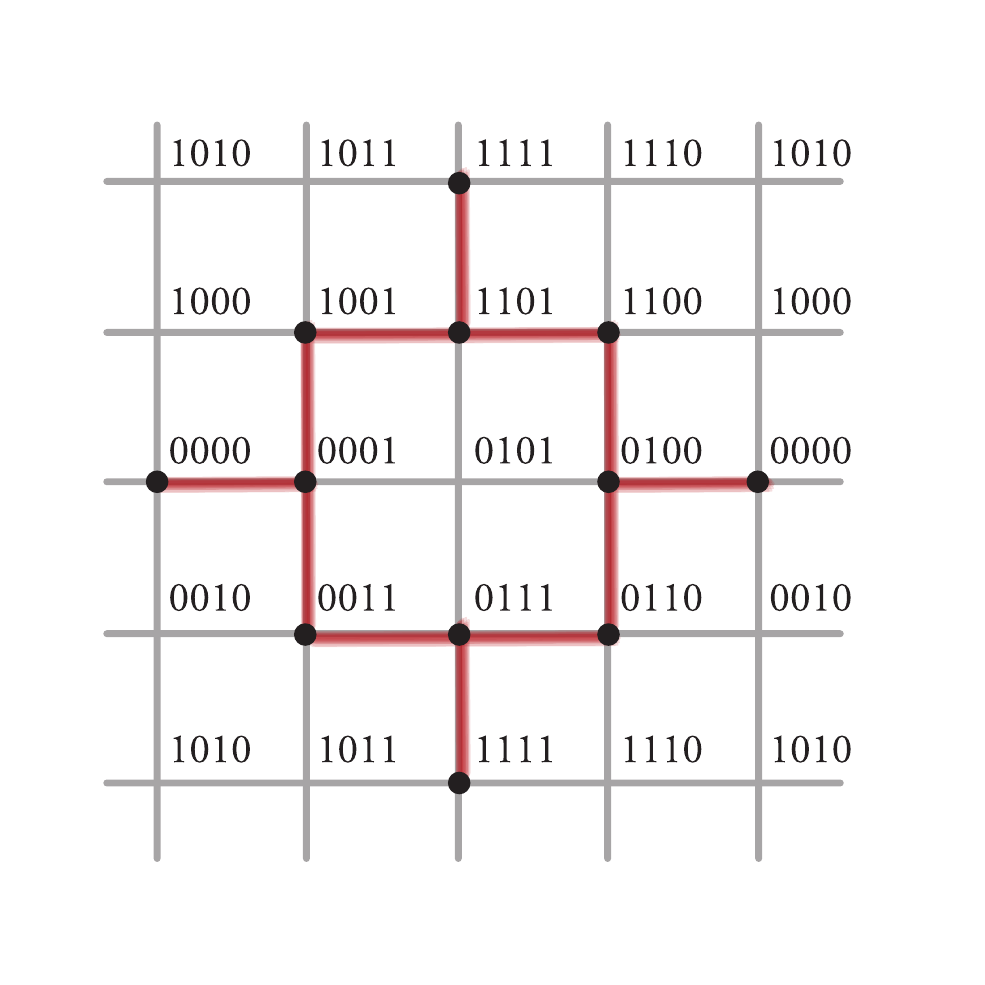}
\caption{Torus embedding}
\label{pic-torus}
\end{figure}

\begin{table}[h!]
\caption{Right side of the TE}
\label{tab-RTE}
\begin{center}
\begin{tabular}{|c|c|c|c|}
\hline
***0 & **1* & *0** & 1***\\
\hline
1*00 & 0*10 & 001* & 100* \\
00*0 & 111* & *000 & 1*11 \\
*110 & *011 & 10*1 & 11*0 \\
\hline
\end{tabular}
\end{center}
\end{table}
The picture \ref{pic-edges} illustrates the table \ref{tab-LTE}.
The picture \ref{pic-torus} visualizes the same subset taking in mind the genus 1 embedding of the 4-cube. 
\begin{Rem}
The essential property of the subgraphs defined by tables \ref{tab-LTE} and \ref{tab-RTE} is that they are exchanged by the indexes transformation $1 \leftrightarrow 4, 2\leftrightarrow 3.$
\end{Rem}

\paragraph{Twisted tetrahedron equation}
The chosen subgraphs properties can be encoded by the weight matrix $W(i,j,k)=W_0\times \exp(t\sum_{u=1}^6 \sigma_{\varphi(u)})$ where the choice of edges is given by the tables above.
We demonstrated in \cite{T1} the following
\begin{Lem}
The matrix $W(i,j+1,k+2)$ gives a weight matrix for the 3D Ising model for each choice of parameters $i,j$ and $k.$ This means that the partition function $Z(t)$ can be obtained as a product
\bea
Z(t)=\prod_{(\alpha,\beta,\gamma)\in\Lambda^*} W_{\alpha\beta\gamma}(i,j,k)
\eea
over the dual lattice $\Lambda^*.$
\end{Lem}

\begin{Th}
\label{th-1}
The weight matrix $W(i,j,k)$ satisfies the equation
\bea
&&W_{653}^A(1,2,3) W_{642}^A(1,2,4) W_{541}^A(1,3,4) W_{321}^A(2,3,4)\nn\\
&&=
W_{356}(2,3,4) W_{246}(1,3,4) W_{145}(1,2,4) W_{123}(1,2,3)
\label{TTE}
\eea
where the upper index $A$ means the conjugation with a linear operator $A$ in each tensor component symbolizing the basis change in $V_f$ exchanging the 1-edges of a 2-face as follows $i_1 \leftrightarrow o_2, i_2\leftrightarrow o_1.$
\end{Th}
\begin{Rem}
We call the equation \ref{TTE} the {\bf twisted tetrahedron equation}.
The importance of the result of theorem \ref{th-1} is that the equation \ref{TTE} is closely resembles the Zamolodchikov tetrahedron equation \cite{Zam} by the following transformation:
\bea
&&{\color{red}P_{16} P_{25}} W_{123}^{{\color{red}A}}(1,2,3) W_{145}^{{\color{red}A}}(1,2,4) W_{246}^{{\color{red}A}}(1,3,4) W_{356}^{{\color{red}A}}(2,3,4){\color{red}P_{16} P_{25}} \nn\\
&&=
W_{356}(2,3,4) W_{246}(1,3,4) W_{145}(1,2,4) W_{123}(1,2,3).\nn
\label{TTE2}
\eea
where $P_{ij}$ are just transpositions in the pair of $i$-th and $j$-th spaces.

The Zamolodchikov tetrahedron equation manifests the integrable properties of the 3D statistical model in virtue of the main result of \cite{T2}. This results is aimed to construct a commutative family including the transfermatrix of the model further targeting in application of the Bethe ansatz technique in an analogous way to the 2-dimensional integrable statistical models \cite{Bel}.
\end{Rem}

\section{Hopfield vertex model}
\subsection{Diagonal combinatorics}
We reproduce the scheme of \cite{T1} in the context of the Ising model on the lattice of figure \ref{edges}.
Let us attach a spin variable to each diagonal by the same rule as for the edge-spins - by the product of the vertex-spins in its ends. 
\begin{figure}[h!]
\center
\includegraphics[width=70mm]{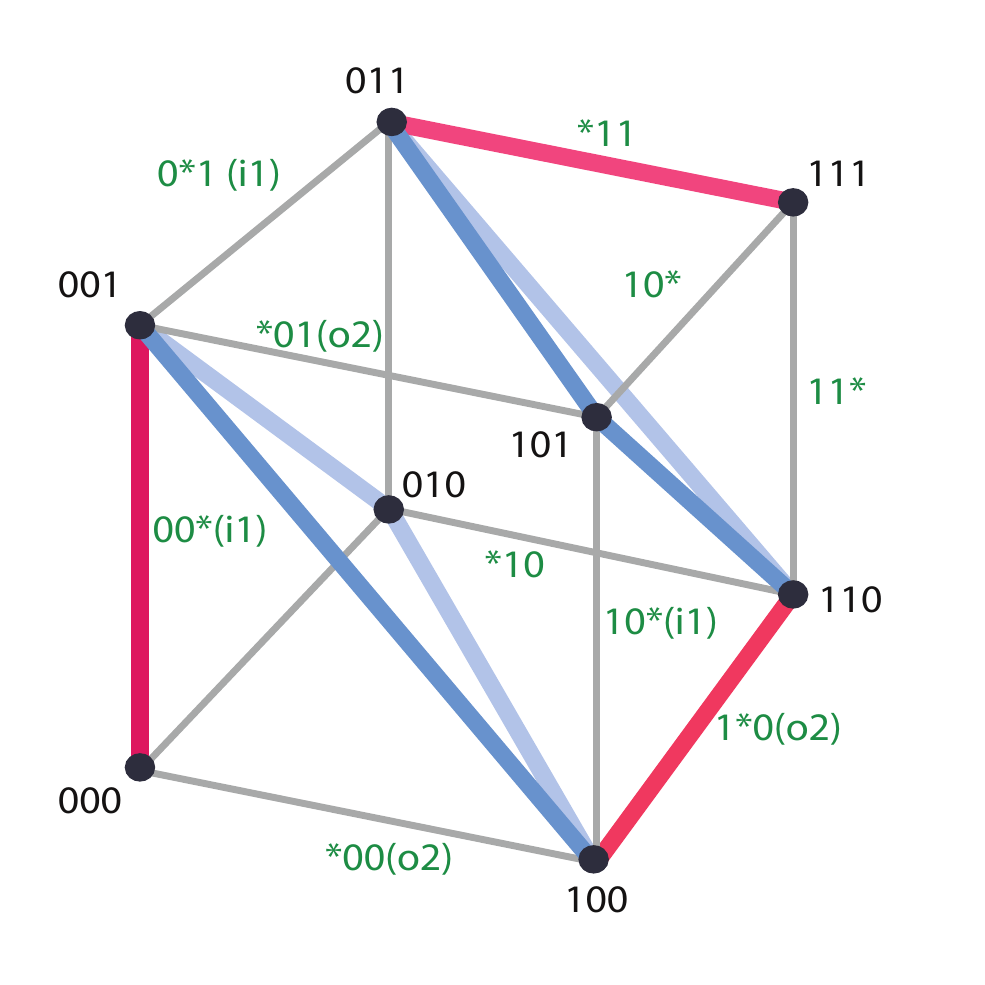}
\caption{Diagonal choice}
\label{fig-diagonals}
\end{figure}
\begin{Lem}
For each 2-face of a 3-cube the value of the diagonal spin is related with the edge-spins by the formula
\bea
s=s_{i_1}\times s_{o_2}=s_{i_2}\times s_{o_1}.\nn
\eea
\end{Lem}
The demonstration is via the figure \ref{fig-diagonals}.

\begin{figure}[h!]
\center
\includegraphics[width=80mm]{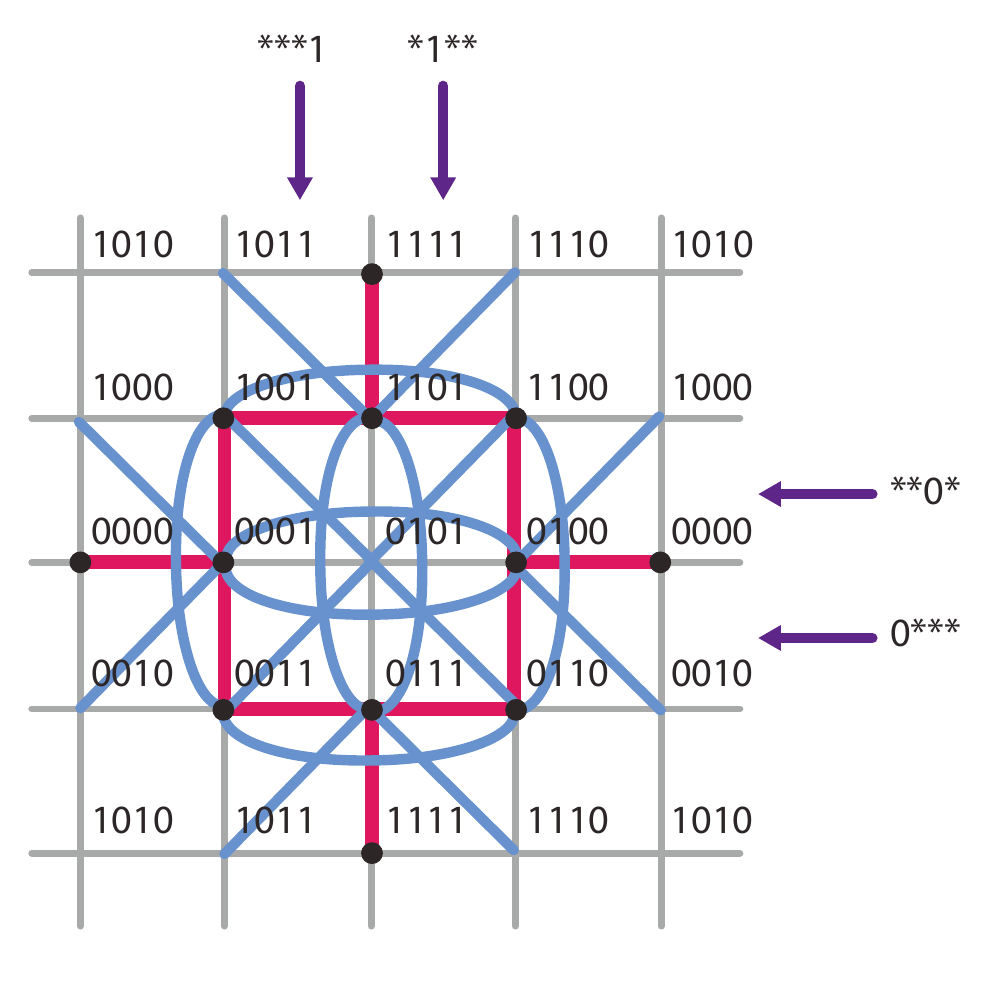}
\caption{Diagonals on LHS}
\label{fig-diag1}
\end{figure}

\begin{figure}[h!]
\center
\includegraphics[width=80mm]{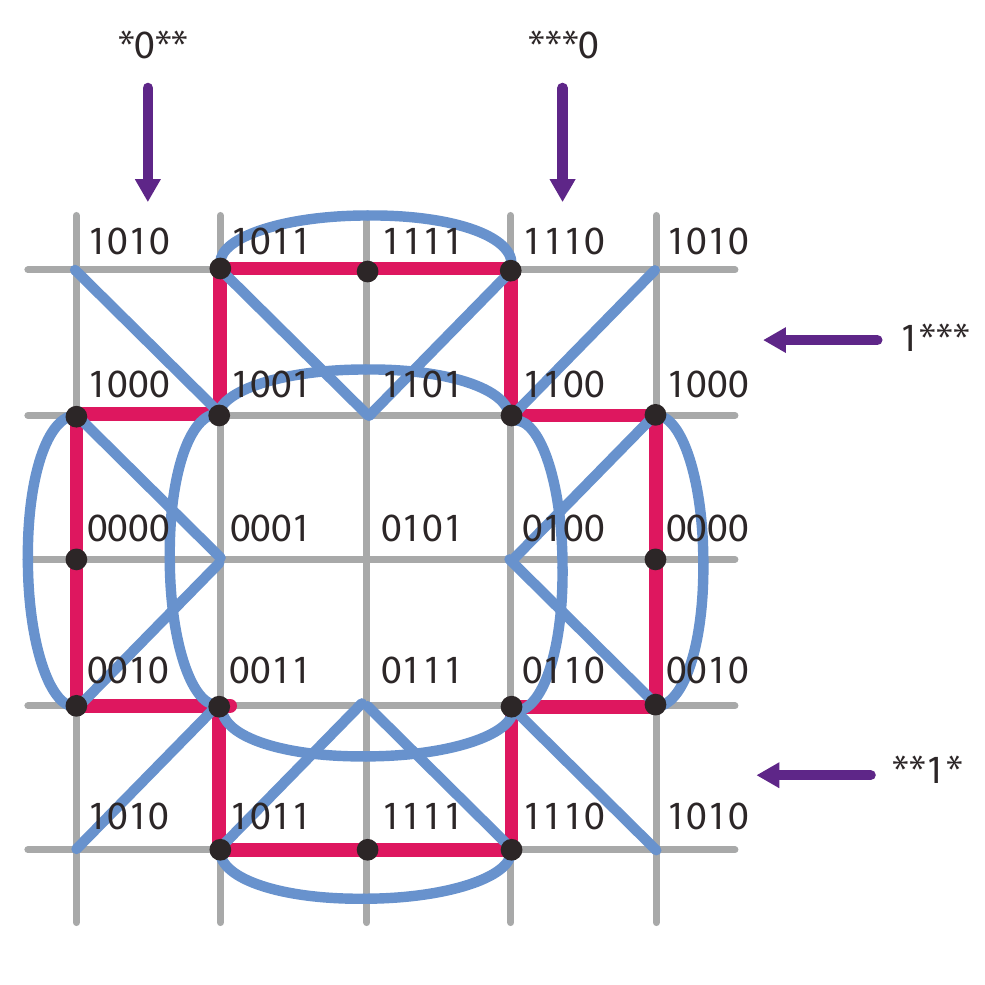}
\caption{Diagonals on RHS}
\label{fig-diag2}
\end{figure}

The choice of the diagonals on the left and right hand sides of the tetrahedron equation are given by the figures \ref{fig-diag1} and \ref{fig-diag2}.

On both figures we singled out the chosen diagonals. We should comment the meaning of these figures. The 3-cubes of the 4-cube are realized as the horizontal and vertical bands. They are marked by the appropriate codes. We denoted the diagonals which are not embedded into the torus by curved blue lines. We need also to note that some of curved lines and embedded diagonals are taken twice.

\begin{Lem}
The chosen diagonals on the left and right hand sides are exchanged by the indexes transformation $1\leftrightarrow 4,~2\leftrightarrow 3.$
\end{Lem}
The demonstration is straightforward.

\subsection{TTE}
\begin{Def}
Let us define the weight matrix $W_h(i,j,k)$ which is constructed with a help of the 3D Ising model weight matrix $W(i,j,k)$ as follows
\bea
W_h(i,j,k)=W(i,j,k)\times \exp(\gamma\sum_f \sigma_{i_1} \sigma_{o_2}).
\eea
Here the sum is over 2-faces of a 3-cube, and the spins variables correspond to the edges of a 2-face.
\end{Def}

\begin{Th}\label{th-2}
The weight matrix $W_h(i,j,k)$ satisfies the twisted tetrahedron equation
\bea
&&W_{653}^A(1,2,3) W_{642}^A(1,2,4) W_{541}^A(1,3,4) W_{321}^A(2,3,4)\nn\\
&&=
W_{356}(2,3,4) W_{246}(1,3,4) W_{145}(1,2,4) W_{123}(1,2,3).
\label{TTE1}
\eea
where the upper index $A$ has the same meaning as in theorem \ref{th-1}.
\end{Th}
The proof  reproduces the proof of theorem 1 of \cite{T1}. It ensues the same observation that the configuration of chosen graph on the left hand side transforms to the graph on the right hand side by the index transformation $1\leftrightarrow 4,~2\leftrightarrow 3$ which can be encoded by the specified matrix notations.

\subsection{Conclusion}
Our main hope concerns the possibility of the Bethe ansatz method application to the description of the critical behavior of the Hopfield neural network. This could be fruitful in such a technique as the simulated annealing in neural networks \cite{An}.


\begin{thebibliography}{50}

\bibitem{T1}
Dmitry V. Talalaev, {\em Towards integrable structure in 3d Ising model}, arXiv:1805.04138

\bibitem{Hop}
W. A. Little.{\em The Existence of Persistent States in the Brain,} Math. Biosciences
 {\bf 19},101-120 (1974) 101

J. J. Hopfield. {\em Neural networks and physical systems with emergent
collective computational abilities.} in Pk. Nat. Academy
Sci., USA, vol. 19, 1982, pp. 2554-2558.

\bibitem{Tso}
S. Recanatesi, M. Katkov, S. Romani and M. Tsodyks. {\em Neural network model of memory retrieval.} Front Comput Neurosci. 2015 {\bf 9}:149. 

S. Romani and M. Tsodyks. {\em Short-term plasticity based network model of place cells dynamics.} Hippocampus 2015 {\bf 25}:94-105. 

Sompolinsky H., Kanter I. (1986). {\em Temporal association in asymmetric neural networks.} Phys. Rev. Lett. {\bf 57}, 2861–2864. 10.1103/PhysRevLett.57.2861

Muraviev IP, Husek D, Frolov AA.;Informational Capacity and Recall Quality in Sparsely Encoded Hopfield-like Neural Network: Analytical Approaches and Computer Simulation.;"Neural Netw. 1997 Jul;10(5):845-855.";12662874 

\bibitem{NP}
Sorin Istrail, {\em Statistical Mechanics, Three-Dimensionality and
NP-completeness I. Universality of Intractability for the Partition Function
of the Ising Model Across Non-Planar Lattices} Conference Paper, January 2000,
DOI: 10.1145/335305.335316

\bibitem{KST} 
 I.G. Korepanov, G.I. Sharygin, D.V. Talalaev, {\em Cohomologies of n-simplex relations,} Mathematical Proceedings of the Cambridge Philosophical Society. 2016. Vol. 161, no. 2. Pp. 203-222.
 
\bibitem{Zam}
Zamolodchikov A.B., {\em Tetrahedra equations and integrable systems in three-dimensional space}, Soviet Phys. JETP {\bf 52} (1980), 325–336.

\bibitem{T2}
Dmitry V. Talalaev, {\em Zamolodchikov Tetrahedral Equation and Higher Hamiltonians of 2d Quantum Integrable Systems}, SIGMA, {\bf 13} (2017), 031, 14 pp. 

\bibitem{Bel}
A.A. Belavin, A.G. Kulakov, R.A. Usmanov, {\em Lectures on theoretical physics.} MCCME, 2001

\bibitem{An}
Da, Y.; Xiurun, G. {\em An improved PSO-based ANN with simulated annealing technique}. Neurocomputing {\bf 63} (2005) 527–533

 
\end{thebibliography}
\end{document}